\documentclass{cjaa}                   

\usepackage{graphicx}                   

\begin{document}

   \title{Detection of Giant Pulses in Pulsar PSR J1752+2359
         }

   \volnopage{Vol.0 (200x) No.0, 000--000}      
   \setcounter{page}{1}          

   \author{A.A. Ershov
      \inst{}\mailto{ershov@prao.psn.ru}
   \and A.D. Kuzmin
      \inst{}
          }
   \offprints{A.A. Ershov}                   

   \institute{Pushchino Radio Astronomy Observatory, Astro Space Center,
      Lebedev Physical Institute, Russian Academy of Sciences,
      Pushchino, 142290, Russia\\
             \email{ershov@prao.psn.ru}
             }

   \date{Received~~2005 November day; accepted~~2005~~month day}

   \abstract{
We report the detection of Giant Pulses (GPs) in the pulsar PSR J1752+2359.
The energy of the strongest GP exceeds the energy of the average pulse by a
factor of 200, in which it stands out from all known pulsars with GPs. PSR
J1752+2359 as well as the previously detected PSR B0031--07 and PSR B1112+50,
belongs to the first group of pulsars found to have GPs without a high magnetic
field at the light cylinder.
            }

   \authorrunning{A. A. Ershov \& A. D. Kuzmin }            
   \titlerunning{Detection of Giant Pulses in Pulsar PSR J1752+2359 }  

   \maketitle

%
%
\section{Introduction}           
\label{sect:intro}
Giant pulses (GPs) are short duration burst-like increases of an
intensity of individual pulses from pulsars. The peak intensities
and energies of GPs greatly exceed the peak intensity and energy
of the average pulse (AP). The energy distribution of GPs has a
power-law. The GPs are much narrower than the AP and their phases
are of stable placement within the AP.

This rare phenomenon was first detected in the Crab pulsar
(Staelin \& Sutton, 1970) and the millisecond pulsar PSR B1937+21
(Wolszczan et al. 1984), both with very strong magnetic fields on
the light cylinder of $B_{\rm LC} = 10^4 - 10^5$\,G. This gave
rise to the suggestion that GPs occur in pulsars with very strong
magnetic fields on the light cylinder and a search of GPs was
oriented on those pulsars. As a result GPs were detected in five
other pulsars with very strong magnetic fields on the light
cylinder: PSR B0218+42 (Joshi et al. 2004), PSR B0540--69
(Johnston \& Romani 2003), PSR B1821--24 (Romani \& Johnston
2001), PSR J1823--3021 (Knight et al. 2005), PSR B1957+20 (Joshi
et al. 2004).

Here we report the detection of GPs in the pulsar PSR J1752+2359.
Correlating this with our previously published data on PSR
B1112+50 (Ershov \& Kuzmin 2003) and PSR B0031--07 (Kuzmin et al.
2004; Kuzmin \& Ershov 2004), we have revealed that GPs exist in
pulsars with relatively low magnetic fields at the light cylinder.


\section{Observations}
\label{sect:Obs}
Observations were performed with the Large Phase Array (BSA) Radio
Telescope at Pushchino Radio Astronomy Observatory of Lebedev
Physical Institute at the frequency of 111\,MHz. This is the
transit telescope with the effective area of about 15\,000 square
meters. One linear polarization was received. We used a
128-channel receiver with the channel bandwidth 20\,kHz. The
sampling interval was 2.56\,ms and the receiver time constant was 3\,ms.
The duration of each observation session was about 3\,min (420
pulsar periods). A total of 120 observations containing 50\,400
pulsar periods were carried out in the mode of recording single
pulses. During the off-line data reduction the signal records were
cleaned of radio interferences and  the inter-channel dispersion
delays were removed. Verification that GPs belong to PSR
J1752+2359 was checked by timing and inter-channels dispersion
delay.

\section{Results}
\label{sect:data}

Figure~1 shows an example of one observation session of GPs. A
group of three bright pulses standing out of the noise background
and underlying weak pulses were observed inside 420 pulsar
periods. The 187 pulses (1 pulse for 270 observed periods) with
$S/N \geq 5$ were selected and analyzed. The observed peak flux
density exceeded the peak flux density of the AP by more than a
factor of 40.
%
\begin{figure}[h]
  \begin{minipage}[t]{0.5\linewidth}
  \centering
  \includegraphics[width=65mm,height=65mm]{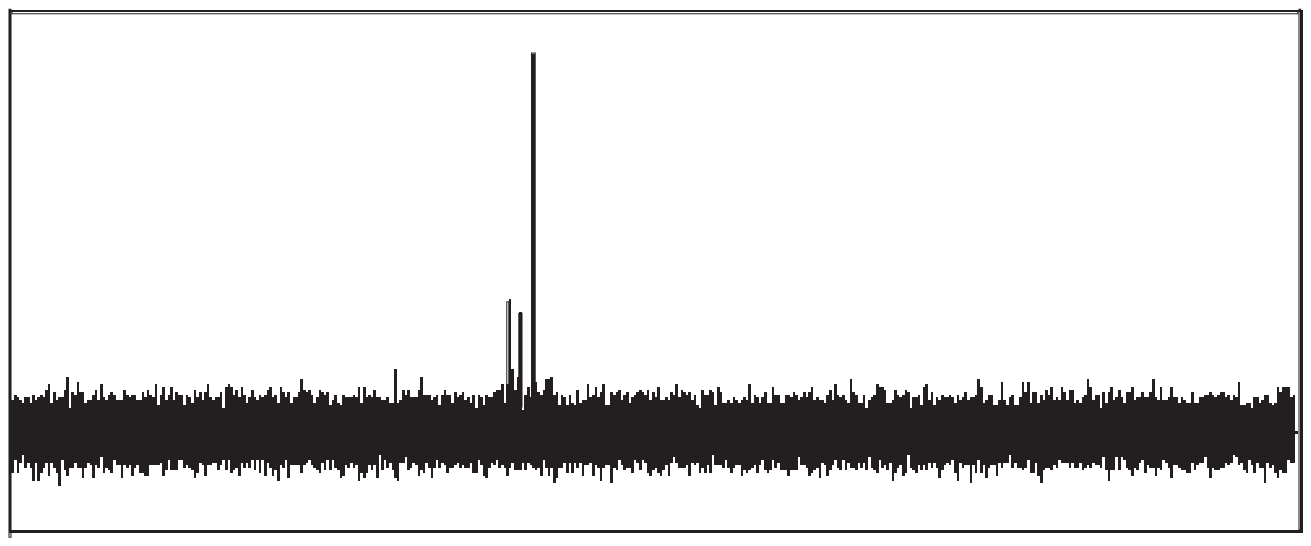}
  \vspace{-5mm}
  \caption{{\small One observation session of GPs. Three large pulses stand
    out of the noise  background and weak pulses are observed inside 420
    pulsar periods.
          }}
  \end{minipage}%
  \begin{minipage}[t]{0.5\textwidth}
  \centering
  \includegraphics[width=65mm,height=65mm]{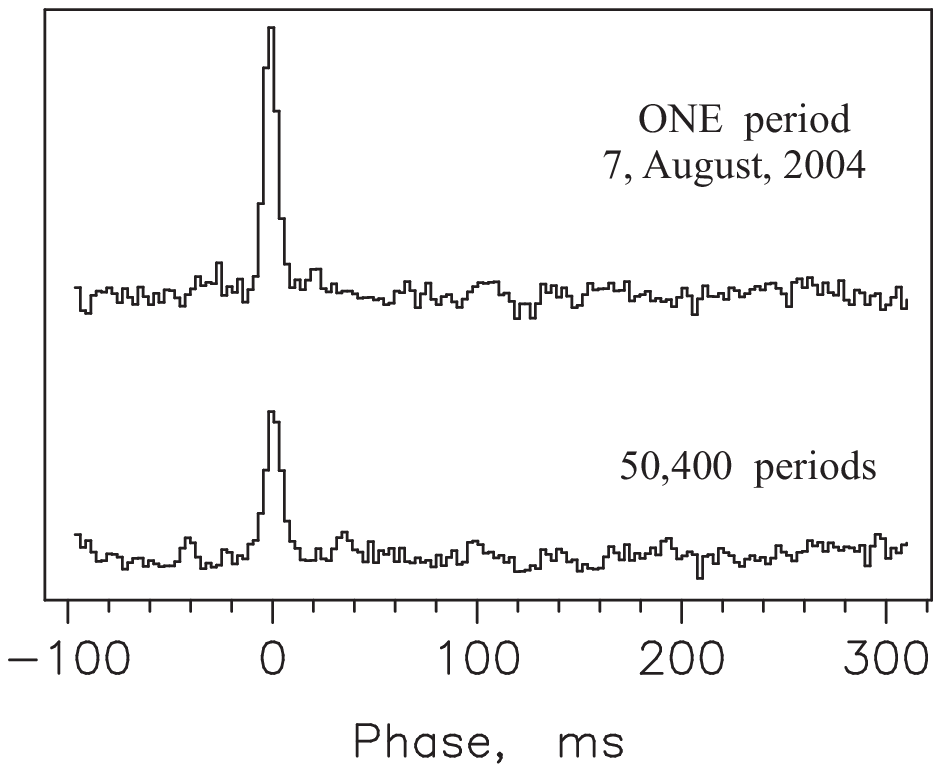}
  \vspace{-5mm}
  \caption{{\small \textbf{(Top)} The strongest observed pulse.
     \textbf{(Bottom)} The average pulse profile containing 50\,400 pulsar
     periods. The intensity of the profiles is shown in arbitrary units.
          }}
  \end{minipage}%
  \label{Fig:fig1-2}
\end{figure}

Figure~2 shows the strongest observed GP together with the AP
averaged over 50\,400 pulsar periods. The peak flux density of the
strongest GP is 105 Jy that exceeds the peak flux density of the
AP by factor of 260. The energy of the strongest observed GP is
920\,Jy\,ms, that exceeds the energy of the AP by factor of
200. This is the most pronounced energy increase factor among the
known pulsars with GPs.

A pulse whose energy exceeded an energy of the average pulse (AP)
by more than a factor of 100 is encountered approximately once in
3000 observed periods.

Along with a large intensity, the distinguishing characteristic of
the previously known pulsars with GPs is for their two-mode pulse
intensity distribution.  At low intensities, the pulse strength
distribution is Gaussian one, but above the certain threshold the
pulse strength distribution is roughly power-law distributed.

In Figure~3 we show the measured cumulative distribution of the
ratio of the observed GPs energy to the AP energy for all pulses
that we have selected and analyzed. The distribution shows a
power-law dependence with index $\alpha = -3.0~\pm~0.4$ (the solid line).
The dotted line represents the possible version of the Gaussian
distribution.

The intrinsic fine structure of GPs is masked by dispersion pulse
broadening of 4.4~ms and receiver time constant of 3\,ms. Therefore
we performed additional observations with higher temporal
resolution. We used a 128-channel receiver with a channel
bandwidth of 1.25\,kHz, sampling interval of 0.81\,ms, and time
constant of 1\,ms. In this mode, we performed 14 observation
sessions containing 6800 pulsar periods.

%
\begin{figure}[h]
  \begin{minipage}[t]{0.5\linewidth}
  \centering
  \includegraphics[width=65mm,height=65mm]{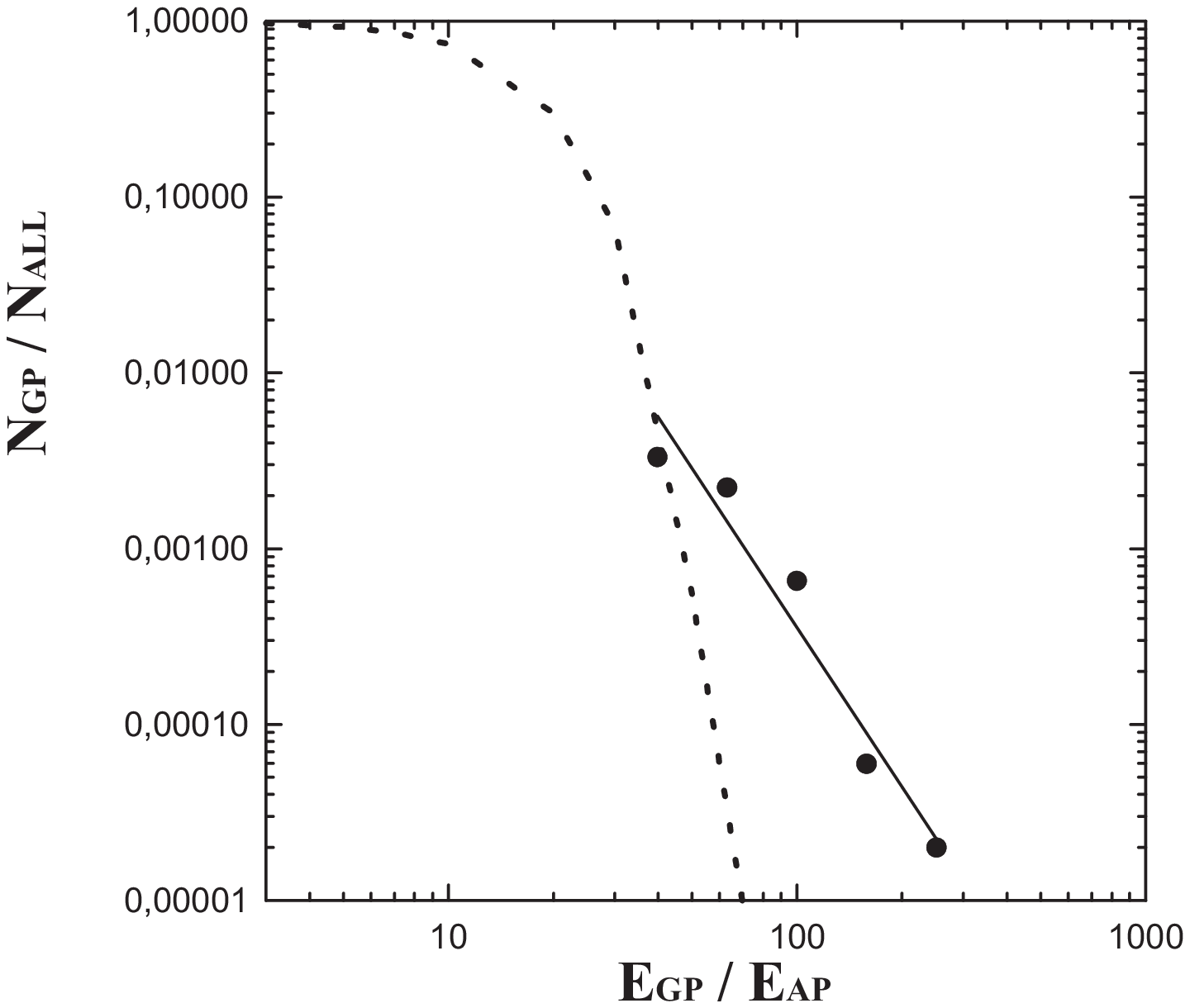}
  \vspace{-5mm}
  \caption{{\small The cumulative distribution of the observed
GP energy $E^{\rm GP}$ as related to the AP energy $E^{\rm AP}$.
The solid line is the observed power-law distribution $N^{\rm
GP}~/~N^{\rm All} \propto (E^{\rm GP}/E^{\rm AP})^{\alpha}$ with
index $\alpha = -3.0~\pm~0.4$. The dotted line represents the
possible version of the Gaussian distribution $N~/~N^{\rm All} =
exp(-a(E~/~E^{\rm AP})^2)$.
          }}
  \end{minipage}%
  \begin{minipage}[t]{0.5\textwidth}
  \centering
  \includegraphics[width=65mm,height=65mm]{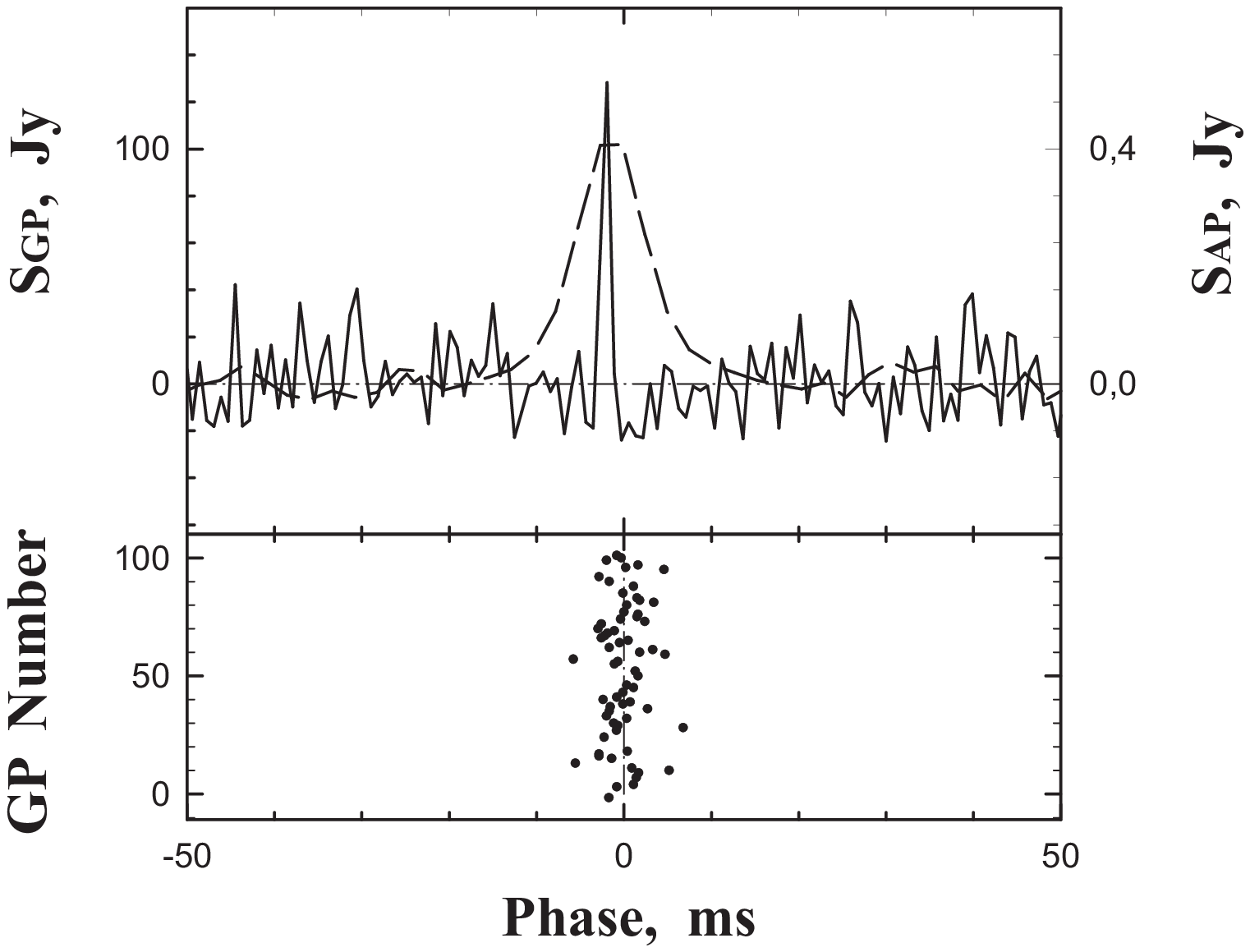}
  \vspace{-5mm}
  \caption{{\small \textbf{(Top)} The  observed high resolution GP
      (bold line) and the AP (dotted line). The observed peak flux density
      of this GP exceeds the peak flux density of the AP by a factor of 320.
      The plot of the AP is presented on a 250 times larger
      scale and flux densities of the observed GP and AP
      are shown separately on the left and right sides of the "y"-axis.
       \textbf{(Bottom)} The phases of the GPs.
          }}
  \end{minipage}%
  \label{Fig:fig3-4}
\end{figure}

Figure~4 (top) shows the high resolution GP (solid line) together
with the AP (dotted line). The observed peak flux density of this
GP exceeds the peak flux density of the AP by factor of 320. The
plot of the AP is presented on a 250 times larger scale, and flux
densities of the observed GP and AP are shown separately, on the
left and right sides of the "y"-axis.The observed width of the GPs
is 1\,ms, which is narrower than the AP by a factor of about 10.

Figure~4 (bottom) shows the phases of the GPs. Giant pulses cluster in a
narrow phase window near the middle of the AP. The clustering is
closer for stronger GPs.The brightness temperature of the
strongest GP is $T_{\rm B} \geq 2 \times 10^{28}$ K.

\section{Discussion}
\label{sect:discussion}
The GPs that we detected in PSR J1752+2359 exhibit all
characteristic features of the classical GPs in PSR B0531+21 and
PSR B1937+21. The peak intensities of the GPs exceed the peak
intensity of the AP by more than a factor of 50. The histograms of
the flux density have a power-law distribution. The GPs are much
narrower than the AP and their phases are stable inside the
integrated profile.

The most important aspect of this report is the fact, that PSR
J1752+2359 (as well as PSR B0031--07 and PSR B1112+50) represents
pulsar with relatively low magnetic field $B_{\rm LC}$ at the
light cylinder. This is in contrast with the canonical suggestion
that GPs occur in pulsar with strong magnetic field at the light
cylinder (e.g. Romani \& Johnston 2001). The detection and first
searches of GPs were performed in pulsars with extremely high
magnetic field at the light cylinder of $B_{\rm LC} = 10^4 -
10^5$~G. Then it was suggested, that GPs originate near the light
cylinder (Istomin 2004). However, detection of GPs in the pulsars
PSR B1112+50 (Ershov \& Kuzmin 2003), PSR B0031--07 (Kuzmin et al.
2004; Kuzmin \& Ershov 2004) and presently reported GPs in PSR
J1752+2359 have revealed that GPs exist also in pulsars with
ordinary magnetic field at the light cylinder of $B_{\rm LC} = 1 -
100$~G. These GPs may be associated with the inner gap emission
region (Gil \& Melikidze 2004; Petrova 2004).

One should note that the GPs of PSR J1752+2359 with relatively low
magnetic field at the light cylinder was observed at low
frequencies of 111\,MHz, whereas GPs of pulsars with strong
magnetic field at the light cylinder were observed mainly at high
frequencies. It is of much interest to observe GPs of PSR
J1752+2359 at high frequencies, where one can realize the better
temporal resolution.

\section{Conclusions}
\label{sect:conclusion}
The Giant Pulses (GPs) from pulsar PSR J1752+2359 have been
detected. The energy of the GPs exceeds the energy of the average
profile by a factor of up to 200, which stands this pulsar out
among the known pulsars with GPs. Cumulative distribution is fit
by a power-law with index $\alpha= -3.0~\pm~0.4$. PSR J1752+2359 as well
as the previously detected PSR B0031--07 and PSR B1112+50 are the
first pulsars with GPs that do not have a high magnetic field at
the light cylinder.

\begin{acknowledgements}
We wish to thank  V.~V. Ivanova, K.~A. Lapaev and A.~S. Aleksandrov
for assistance during observations.  This work was supported in
part by the Russian Foundation for Basic Research (project No
05-02-16415). We are grateful to Dr. Steve Shore for valuable
comments.

\end{acknowledgements}

\label{lastpage}

\end{document}